\documentclass[12pt]{elsarticle}
\usepackage{graphicx}
\usepackage{amsmath}
\usepackage{subfig}

\newcommand{\N}{{\rm I}\!{\rm N}}

\begin{document}

	\title{Tornadoes and related damage costs statistical modeling with a semi-Markov approach}
	
	\author{Chiara Corini}  
	\address{Dipartimento di Metodi e Modelli per l'Economia, il Territorio e la Finanza,
		Universit\`a degli studi di Roma La Sapienza, 00161, Rome, Italy }
	\author{Guglielmo D'Amico}  
	\address{Dipartimento di Farmacia, 
		Universit\`a `G. D'Annunzio' di Chieti-Pescara,  66013 Chieti, Italy}
	\author{Raimondo Manca}
	\address{Dipartimento di Metodi e Modelli per l'Economia, il Territorio e la Finanza,
		Universit\`a degli studi di Roma La Sapienza, 00161, Rome, Italy }
	\author{Filippo Petroni}
	\address{Dipartimento di Scienze Economiche ed Aziendali,
		Universit\`a degli studi di Cagliari, 09123 Cagliari, Italy}
	\author{Flavio Prattico}
	\address{Dipartimento di Metodi e Modelli per l'Economia, il Territorio e la Finanza,
		Universit\`a degli studi di Roma La Sapienza, 00161, Rome, Italy }

	\begin{abstract}
		We propose a statistical approach to tornadoes modeling for predicting and simulating occurrences of tornadoes and accumulated cost distributions over a time interval.
		This is achieved by modeling the tornadoes intensity, measured with the Fujita scale, as a stochastic process. Since the Fujita scale divides tornadoes intensity into six states, it is possible to model the tornadoes intensity by using Markov and semi-Markov models. We demonstrate that the semi-Markov approach is able to reproduce the duration effect that is detected in tornadoes occurrence. The superiority of the semi-Markov model as compared to the Markov chain model is also affirmed by means of a statistical test of hypothesis. As an application we compute the expected value and the variance of the costs generated by the tornadoes over a given time interval in a given area.
		The paper contributes to the literature by demonstrating that semi-Markov models represent an effective tool for physical analysis of tornadoes as well as for the estimation of the economic damages to human things. 
		
	\end{abstract}
	
	
	\maketitle

	\section{Introduction}
	
	Every year tornadoes cause deaths and several damages to people and things. Only in the USA, tornadoes killed in average more than 100 people per year from 2004 to 2013. Just to give and example of the monetary damages of tornadoes in the USA, in 2013 they cost about 200 millions of dollars  \cite{simmons2013normalized}. In this scenario, the development of techniques to estimate and model the probabilities of these events is needed and can be of great benefit for the society. Many researchers are working on this subject, see e.g.  \cite{sisson2006case,de2009assessing,obeysekera2011climate}. The approaches used can be typically divided into two main groups, one
	analytical and another statistical (e.g. see \cite{bryan2009evaluation} and \cite{dotzek2003statistical}, respectively).
	
	Here we propose a statistical approach based on semi-Markov model. This kind of models generalize the more common Markov chain models and their main feature is the possibility to reproduce the duration effect of the considered random phenomenon. This is made possible by considering
	sojourn
	times in the states of the process, that are distributed according to any type of probability distribution functions, non-memoryless distributions included. 
	In this work we choose to model the tornado's intensity as a stochastic process. The tornado's intensity is measured by the Fujita scale which is an empirical scale related to the gravity of the damages produced by the tornado. Since the Fujita scale divides tornadoes intensity into six states, it is possible to model the tornadoes intensity by using semi-Markov models. The database used in this work is made available from the National Oceanic and Atmospheric Administration (USA) that counts of more than 60 000 tornadoes from 1950
	until 2013.
	The proposal of a semi-Markov model for modeling tornadoes allows the estimation of probability of an occurrence of a tornadoes with a certain intensity at each time in a given location. This also gives the possibility to compute the total costs of damages caused by the tornadoes which is a relevant indicators of environmental hazards.  
	The paper is organized as follow. In the next Section we introduce the
	database and the object of investigation. In Section 3 we present the semi-Markov model and the related reward (cost) process. Section 4 shows the main application of the model to the tornado process. At last, in Section 5 we give some concluding remarks.

	\section{Database}
\begin{figure}
		\centering
		\includegraphics[height=7cm]{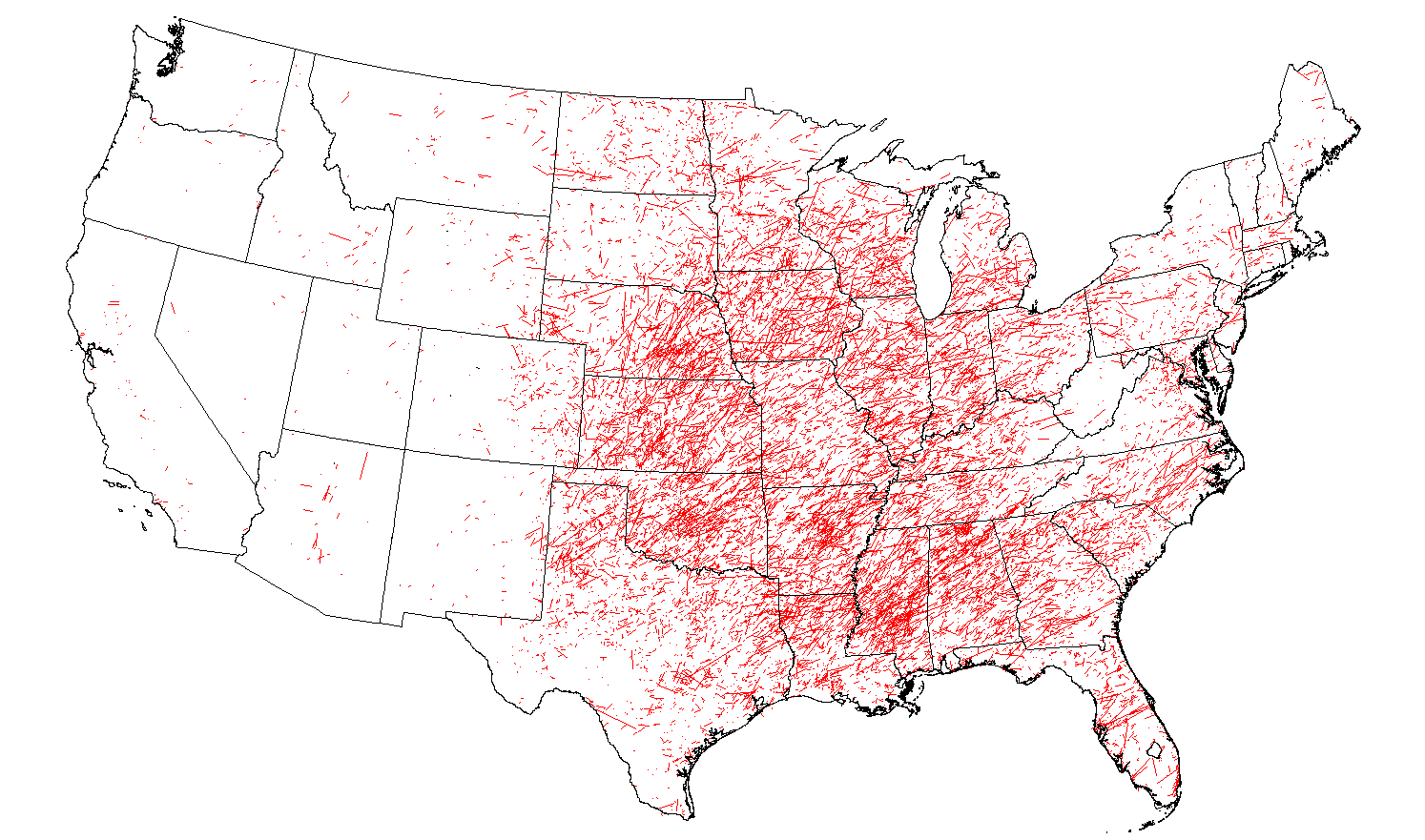}
		\caption{Geographical distribution of the database's events, extrapolated from http://www.spc.noaa.gov/gis/svrgis/images/tornado.png}\label{db}
	\end{figure}

	The data that we use in this work come from the National Oceanic and Atmospheric Administration's (NOAA)  National Weather Service and it are freely available on the website www.spc.noaa.gov/wcm/\#data. In the database are collected almost 60 000 events from 1950 to 2013, all of them geographically distributed in the USA (as it is possible to see in Figure \ref{db}).  
	
	For each event date, time,  state, F-scale, injuries, fatalities, starting latitude and longitude, ending latitude and longitude are recorded. The physical quantity of our interest is the F-scale, (the Fujita scale). This is an empirical scale that measure tornado intensity based on the damage produced to man-made structures. It can be also almost joined to the wind speed, e.g. for a tornado classified F0 the wind speed can goes from 64 to 116 $m/s$, instead for a F5 tornado from 419 to 512 $m/s$ \cite{fujita1973tornadoes}. As it is well known, the Fujita scale admits six values of tornado intensity that go from F0 to F5. As it is a discrete scale, the tornado intensities, measured by the Fujita scale, can be naturally  modeled through semi-Markov models.

	\section{Semi-Markov Process}
	We define an homogeneous semi-Markov process  with values in a finite state space $E=\{1,2,...,m\}$, see for example \cite{limnios2001semi,janssen2006applied}. Let $(\Omega,\mathbf{F},P)$ be a probability space; we consider two sequences of random variables $J=\{J_{n}\}_{n\in \N}$ and $T=\{T_{n}\}_{n\in \N}$ where
	\begin{equation*}
	J_{n}:\Omega\rightarrow E\,;\,\,\,\,\,\,T_{n}:\Omega \rightarrow \N
	\end{equation*}
	\indent They denote the state and the time of the n-th transition of the system, respectively. In our application $J_{n}$ is the intensity of the n-th tornadoes and $T_{n}$ the time of its occurrence.\\
	\indent We assume that $(J,T)$ is a Markov Renewal Process on the state space $E \times \N$ with kernel $Q_{ij}(t),\,\,i,j\in E, t\in \N$. The kernel has the following probabilistic interpretation:
	\begin{equation}
	\label{due}
	\begin{aligned}
	&P[J_{n+1}=j, T_{n+1}-T_{n}\leq t |\sigma(J_{h},T_{h}),\,h\leq n, J_{n}=i]=\\
	& P[J_{n+1}=j, T_{n+1}-T_{n}\leq t |J_{n}=i]=Q_{ij}(t),
	\end{aligned}
	\end{equation}
	where $(\sigma(J_{h},T_{h}),\,h\leq n)$ represents the set of past values of the Markov renewal process $(J,T)$. Relation $(\ref{due})$ asserts that the knowledge of the last tornado's intensity suffices to give the conditional distribution of the couple $(J_{n+1}, T_{n+1}-T_{n})$ whatever the past values of the variables might be.\\ 
	\indent It is simple to realize that $p_{ij} := P[J_{n+1}=j \mid J_{n}=i]= \mathop {\lim }\limits_{t\, \to \,\infty } Q_{ij}(t); \, i, j \in E, \, t\in \N$
	where
	$
	{\bf P} = (p_{ij})
	$
	is the transition probability matrix of the embedded Markov chain $J_{n}$.\\
	\indent Simple probabilistic reasoning allows the computation of the conditional probability distribution of the sojourn time $T_{n+1}-T_{n}$ in the state $J_{n}$ given that next visited state is $J_{n+1}$. In formula:
	\begin{eqnarray}
	\label{quattro}
	&&G_{ij}(t):=P\{T_{n+1}-T_{n}\leq t|J_{n}=i, J_{n+1}=j\}=\nonumber \\ 
	&&\left\{
	\begin{array}{cl}
	\ \frac{Q_{ij}(t)}{p_{ij}}  &\mbox{if $p_{ij}\neq 0$}\\
	1  &\mbox{if $p_{ij}=0$}\\
	\end{array}
	\right.
	\end{eqnarray}
	\indent The $G_{ij}(\cdot)$ denotes the waiting time distribution function in state $i$ given that, with next transition, the process will be in the state $j$. The sojourn time distribution $G_{ij}(\cdot)$ can be any distribution function. We recover the discrete time Markov chain when the $G_{ij}(\cdot)$ are all geometrically  distributed. Therefore we should find out whether the inter-arrival times between two tornadoes of given intensities follows a geometric distribution or not. This is a primary question to which we will respond in next section.\\
	\indent Now it is possible to define the time homogeneous semi-Markov chain $Z(t)$ as
	\begin{equation}
	\label{cinque}
	Z(t)=J_{N(t)},\,\,\,\forall t\in \N
	\end{equation}
	\noindent where $N(t)=\sup\{n\in \N: T_{n}\leq t\}$. Then $Z(t)$ represents the state of the system for each waiting time.\\ 
	\indent At this point we introduce the discrete backward recurrence time process linked to the semi-Markov chain. For each time $t\in \N$ we define the following stochastic process:
	\begin{equation}
	\label{diciannove}
	B(t)=t-T_{N(t)}.
	\end{equation}
	\indent We call it discrete backward recurrence time process. It denotes the time elapsed from the occurrence of the last tornado to the current time $t$.\\
	\indent The joint stochastic process $(Z(t),B(t), t\in \N)$ with values in $E\times \N $ is a Markov process. That is:
	\begin{displaymath}
	\begin{aligned}
	& P[Z(T)\!=\!j, B(T)\! = \!v'|\sigma(Z(h),B(h)),h\! \leq \!t, Z(t)\!=\!i, B(t)\!=\!v]\\
	& =P[Z(T)=j, B(T)= v'|Z(t)=i, B(t)=v]=:\, ^{b}\phi_{ij}^{b}(v;v',t).
	\end{aligned}
	\end{displaymath}
	with the following evolution equation, see e.g. \cite{d2012semi}:
	\begin{equation}
	\label{ventuno}
	\begin{aligned}
	& ^{b}\phi_{ij}^{b}(v;v',t)=\delta_{ij}\frac{[1-\sum_{a\in E}Q_{ia}(t+v)]}{[1-\sum_{a\in E}Q_{ia}(v)]}1_{\{v'=t+v\}}\\
	& +\sum_{k\in E}\sum_{s=1}^{t}\frac{Q_{ik}(s+v)-Q_{ik}(s+v-1)}{[1-\sum_{a\in E}Q_{ia}(v)]}\,^{b}\phi_{kj}^{b}(0;v',t-s).
	\end{aligned}
	\end{equation}
	\indent Expression $(\ref{ventuno})$ provides the probability of having a tornado of intensity $j$ after $t-v'$ periods and no additional tornado within the times $\{t-v'+1, t-v'+2,\ldots,t\}$ given that the last tornado occurred $v$ periods before the present time and was of intensity $i$.\\
	\indent We can now define the accumulated discounted reward (cost), $\xi(t)$, during the time interval $(0,t]$, by the following relation,
	\begin{equation}
	\xi(t)=\sum_{n=1}^{N(t)}  \psi_{J_{n}}\, e^{-\delta T_{n}}
	\label{1}
	\end{equation}
	where $\psi_{J_{n}}$ is the cost caused by the n-th tornado that had an intensity $J_{n}$. This cost has to be discounted using a deterministic force of interest $\delta$ and the time $T_{n}$ of occurrence of the event. The total damage over the time interval $[0,t]$ is obtained by summation over the random number of tornadoes $N(t)$ up to time $t$.\\
	\indent In the application section we will compute the expected value $E[\xi(t)]$ and the second order moment $E[\xi^{2}(t)]$. For an extended treatment of the semi-Markov reward process see e.g. \cite{stenberg2006semi}.
	
	%
	%
	
	\section{Application to real data}
	
	\subsection{Test}
	The first step of our application is to test the validity of the Markov chain hypothesis and to do that we apply a test of hypothesis proposed by \cite{stenberg2006semi} and here shortly described. As already stated, the model can be considered Markovian if the sojourn times are geometrically distributed. The probability distribution function of the sojourn time in state $i$ before making a transition in state $j$ has been denoted by $G_{ij}(\cdot)$. Define the corresponding probability mass function by 
	\begin{eqnarray}
	\label{pmf}
	&&g_{ij}(t)=P\{T_{n+1}-T_{n}= t|J_{n}=i, J_{n+1}=j\}=\nonumber \\ 
	&&\left\{
	\begin{array}{cl}
	\ G_{ij}(t)-G_{ij}(t-1)  &\mbox{if $t > 1$}\\
	G_{ij}(1)  &\mbox{if $t=1$}\\
	\end{array}
	\right.
	\end{eqnarray}
	Under the geometrical hypothesis the equality $g_{ij}(1)(1-g_{ij}(1))-g_{ij}(2)=0$ must hold, then a sufficiently strong deviation from this equality has to be interpreted as an evidence against the Markovian hypothesis and in favor of the semi-Markov model. The test-statistic is the following:
	\begin{equation}
	\label{test}
	\hat{S}_{ij}=\frac{\sqrt{N(i,j)}\big(\hat{g}_{ij}(1)(1-\hat{g}_{ij}(1))-\hat{g}_{ij}(2)\big)}{\sqrt{\hat{g}_{ij}(1)(1-\hat{g}_{ij}(1))^{2}(2-\hat{g}_{ij}(1))}}.
	\end{equation}
	\noindent where $N(i,j)$ denotes the number of transitions from state $i$ to state $j$ observed in the sample and $\hat{g}_{ij}(x)$ is the empirical estimator of the probability $g_{ij}(x)$ which is given by the ratio between the number of transition from $i$ to $j$ occurring exactly after $x$ unit of time and $N(i,j)$. This statistic, under the geometrical hypothesis $H_{0}$ (or Markovian hypothesis), has approximately the standard normal distribution, see \cite{stenberg2006semi}.\\
	\indent We applied this procedure to our data to execute tests at a significance level of $95\%$. Because we have $6$ states we estimated the  $6\times (6-1)$ waiting time distribution functions and for each of them we computed the value of the test-statistic (\ref{test}). The geometric hypothesis is rejected for $17$ of the $30$ distributions. In Table \ref{test1} we show the results of the test applied to the waiting time distribution functions for few states.\\
	\begin{table}
		\label{table1}  
		\begin{center}
			\begin{tabular}{llll}
				\hline\noalign{\smallskip}
				\textbf{state} & \textbf{state} & \textbf{score} & \textbf{decision} \\
				\noalign{\smallskip}\hline\noalign{\smallskip}
				$i=1$ & $j=2$ & 9.79 & $H_{0}$ rejected \\
				$i=1$ & $j=3$ & 4.43 & $H_{0}$ rejected \\
				$i=3$ & $j=1$ & 4.24 & $H_{0}$ rejected \\
				$i=4$ & $j=1$ & 5.50 & $H_{0}$ rejected \\
				\noalign{\smallskip}\hline
			\end{tabular}
			\caption{Results of the Test}
			\label{test1}
		\end{center}
	\end{table}
	\indent The large values of the test statistic suggest the rejection of the Markovian hypothesis in favor of the more general semi-Markov one.\\
	
	\subsection{Probability Transition Matrices}
	
	To set the Markov model and the semi-Markov one, described in previous section, we use the Matlab Application Semi-Markov Toolbox \cite{tool}. This application allows to create Markov and semi-Markov models starting from real discrete data of a given phenomenon. The outputs consist of synthetic time series, of the same length as the real one, generated by means of Monte Carlo simulation and the probability transition matrices. These are practically the core of the models and allow to use them for different purposes, such as time series generation, forecasting and simulation of the phenomenon of interest.  The Monte Carlo algorithm consists in repeated random sampling to compute successive visited states of the random variables $\{J_{0}, J_{1},...\}$  up to the horizon time $L$. The difference of the semi-Markov with respect to Markov is to consider also the jump times $\{T_{0},T_{1},...\}$.
	The algorithm for semi-Markov model consists of 4 steps:\\
	1) Set $n=0$, $J_{0}=i$, $T_{0}=0$, horizon time$=L$;\\
	2) Sample $J$ from $\hat{p}_{J_{n}}$ and set $J_{n+1}=J(\omega)$;\\
	3) Sample $W$ from $\hat{G}_{J_{n},J_{n+1}}$ and set $T_{n+1}=T_{n}+W(\omega)$;\\
	4) If $T_{n+1}\geq L$ stop\\
	\indent else set $n=n+1$ and go to 2).

\begin{figure}
	\centering
	\includegraphics[height=7cm]{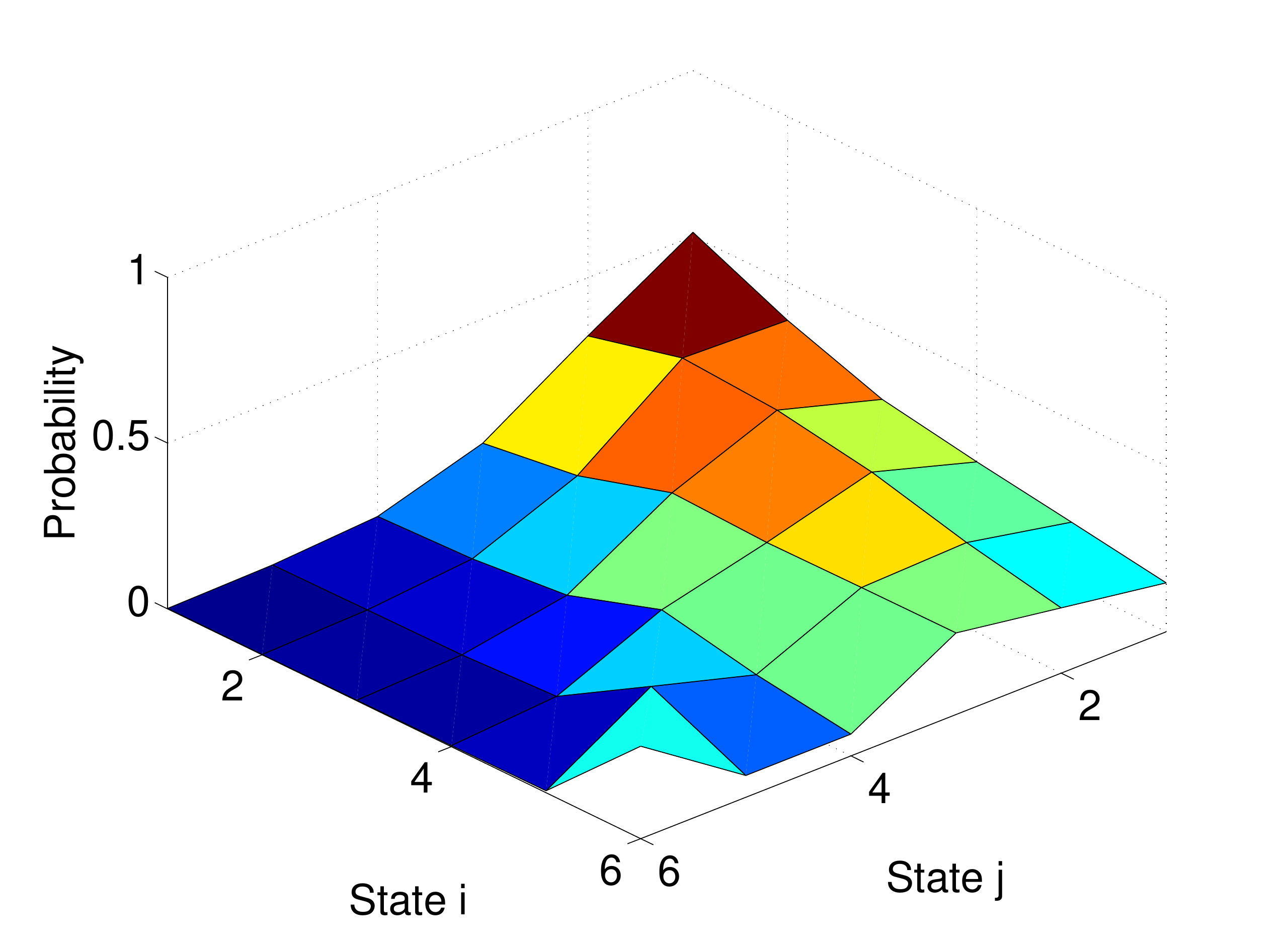}
	\caption{Transition probability matrix of the embedded Markov model.}\label{pm}
\end{figure}

Here below we show the results of the application in terms of transition probability matrices of the two considered models. Particularly, in Figure \ref{pm} we show graphically the transition probability matrix of the embedded Markov model.

\begin{figure}
	\centering
	\includegraphics[height=18cm]{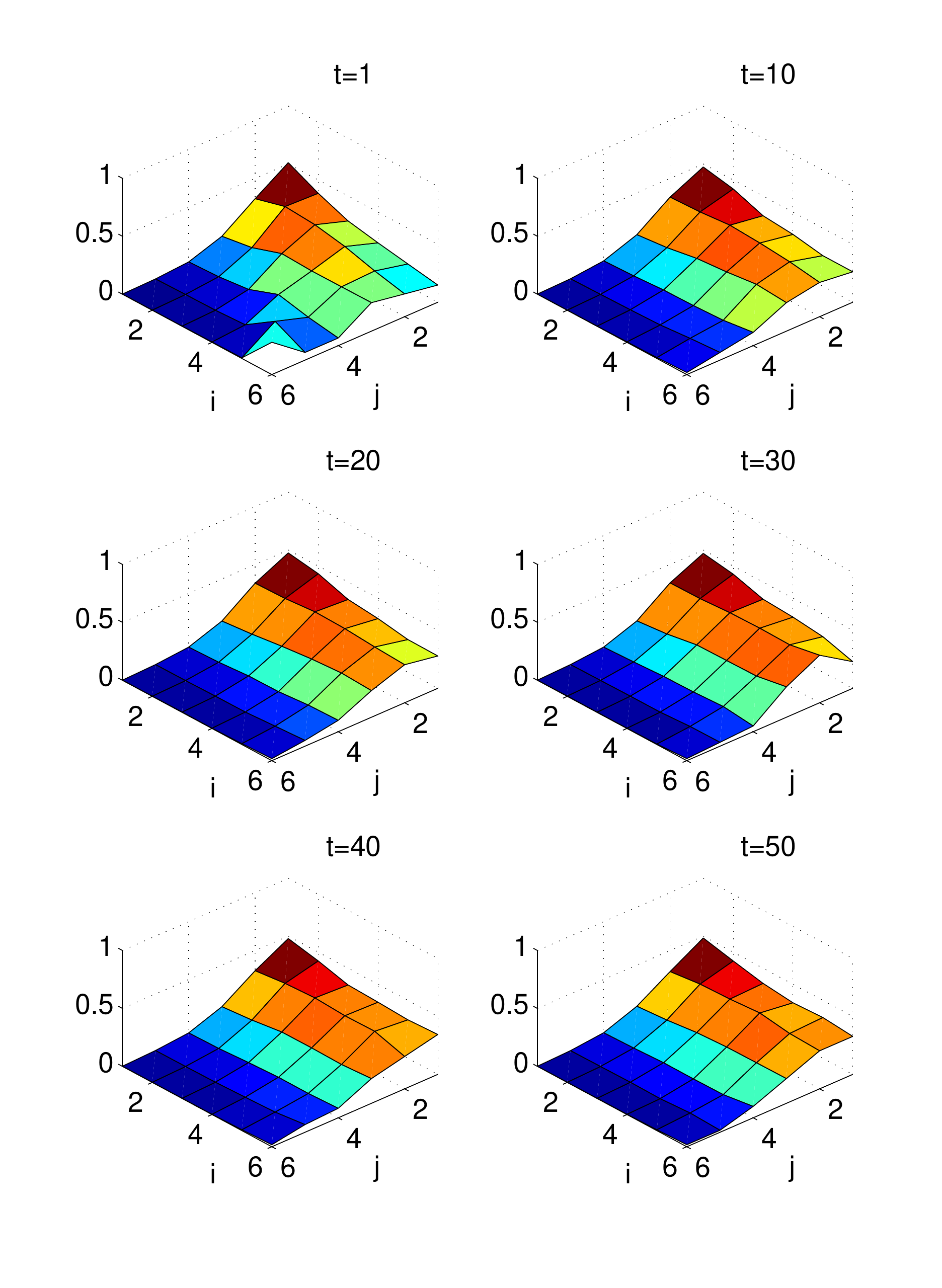}
	\caption{Transition probability matrix of the semi-Markov model varying the time $t$.}\label{dt}
\end{figure}

\begin{figure}
	\centering
	\includegraphics[height=18cm]{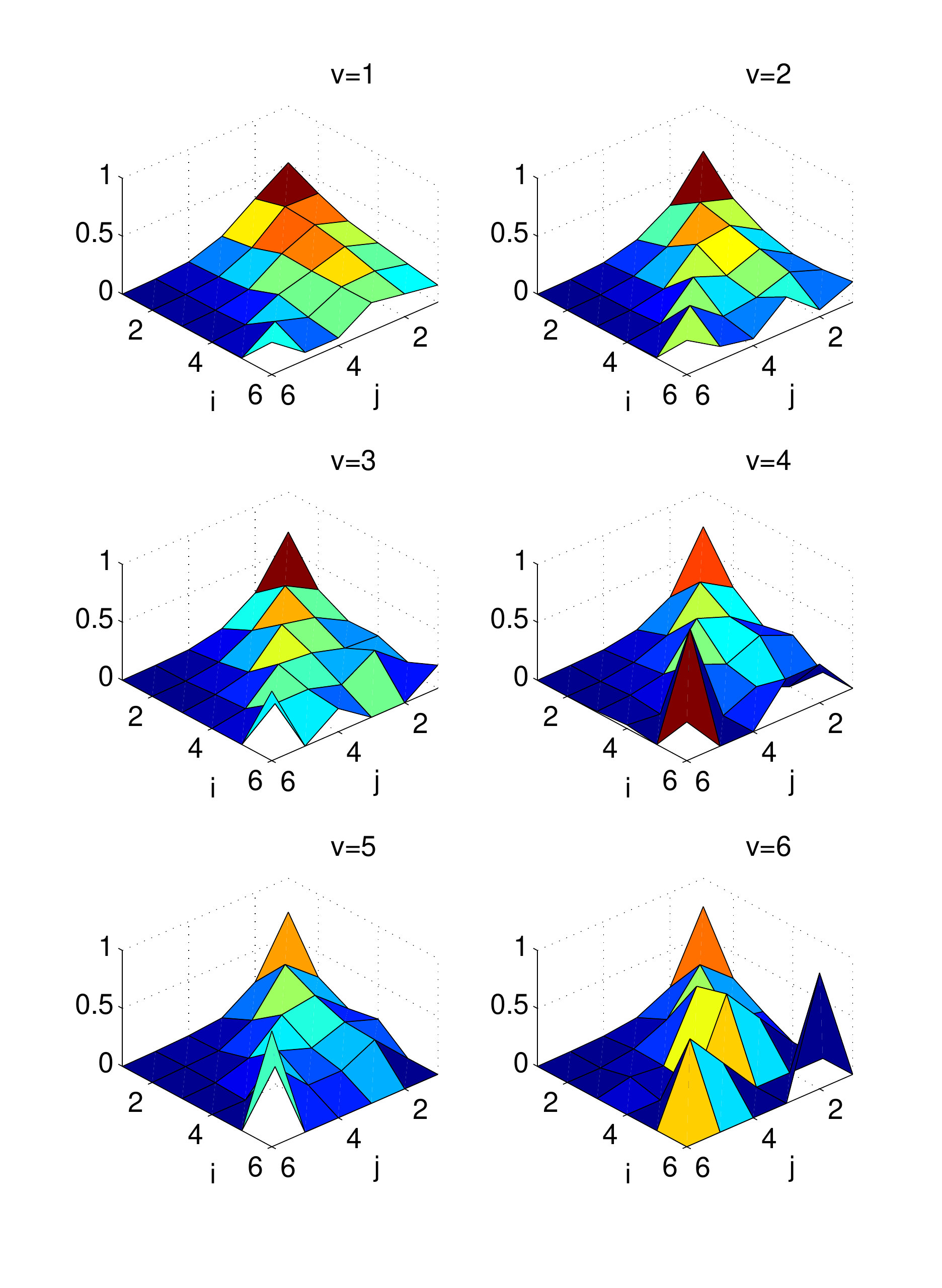}
	\caption{Transition probability matrix of the semi-Markov model varying the backward $v$.}\label{dv}
\end{figure}

In Figures \ref{dt} and \ref{dv} instead, we show the transition probability matrices of the semi-Markov model. The different matrices are plotted by varying the time $t$, by fixing $v=1$, (Figure \ref{dt}) and the backward $v$, by fixing $t=1$ (Figure \ref{dv}). As it is possible to note the dependence of the tornado process by the backward is more strong with respect to the time. This is evident in Figure \ref{dv}, where for little variations of the backward we have great variation on the probability transition matrices. From Figure \ref{dv} we can continue to highlight the great dependence of the process by the backward from the observation of the extreme states. For example if we have an F5 tornado (state 6), we can observe that the probability to have, in the next step, a tornado with the same intensity increase with the increasing of the backward. A similar observation can be made for the virtual transition on the state $1$, that corresponds to F0 intensity. More generally we can note, at the increasing of the backward,  a movement of mass on the main diagonal of the transition probability matrices.

\subsection{Reward application}

\begin{figure}
	\centering
	\includegraphics[height=8cm]{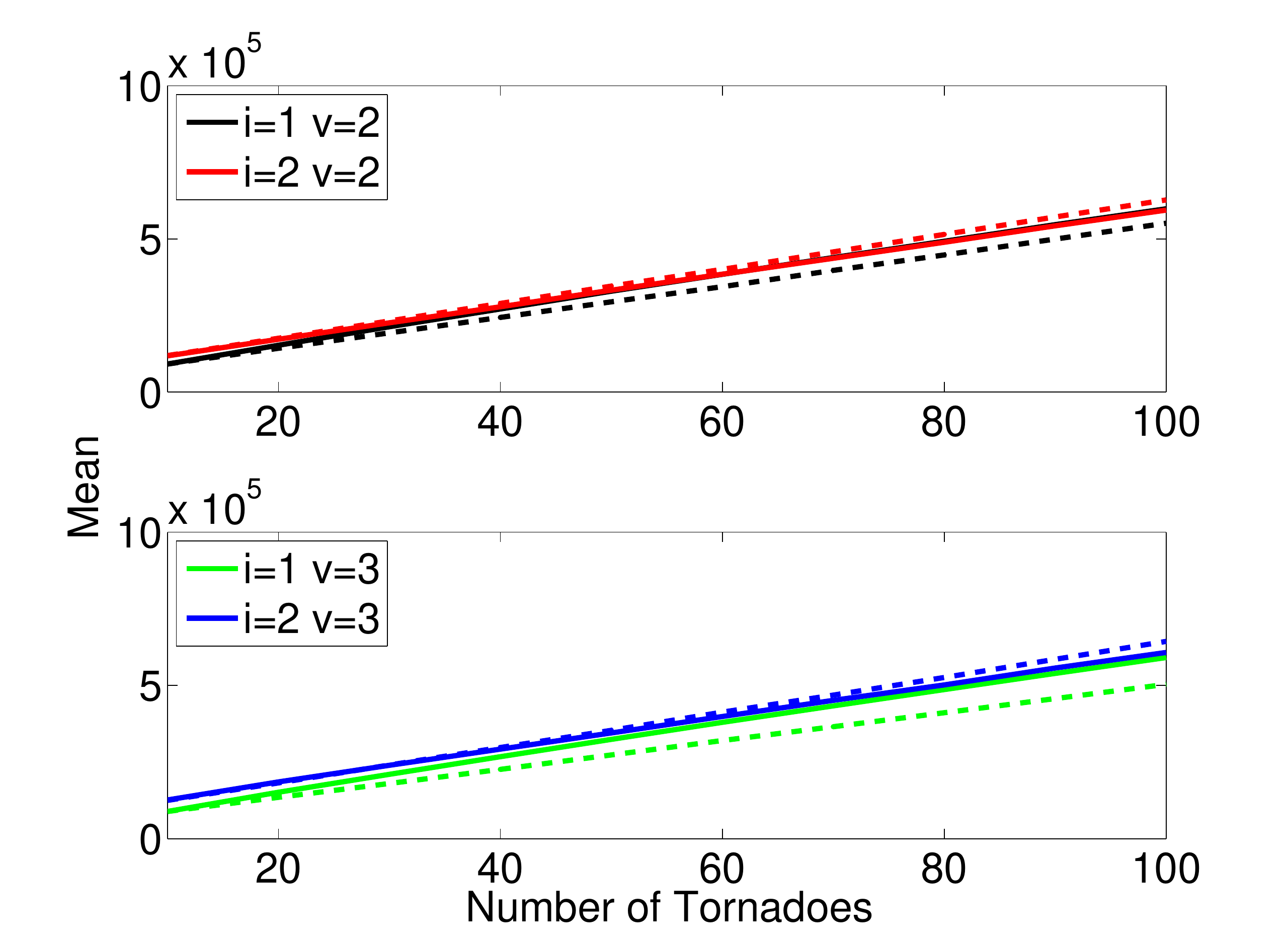}
	\caption{Expected value of the accumulated discounted reward. Comparison between real (continuous line) and synthetic data (dashed line).}
	\label{mean}
\end{figure}

\begin{figure}
	\centering
	\includegraphics[height=8cm]{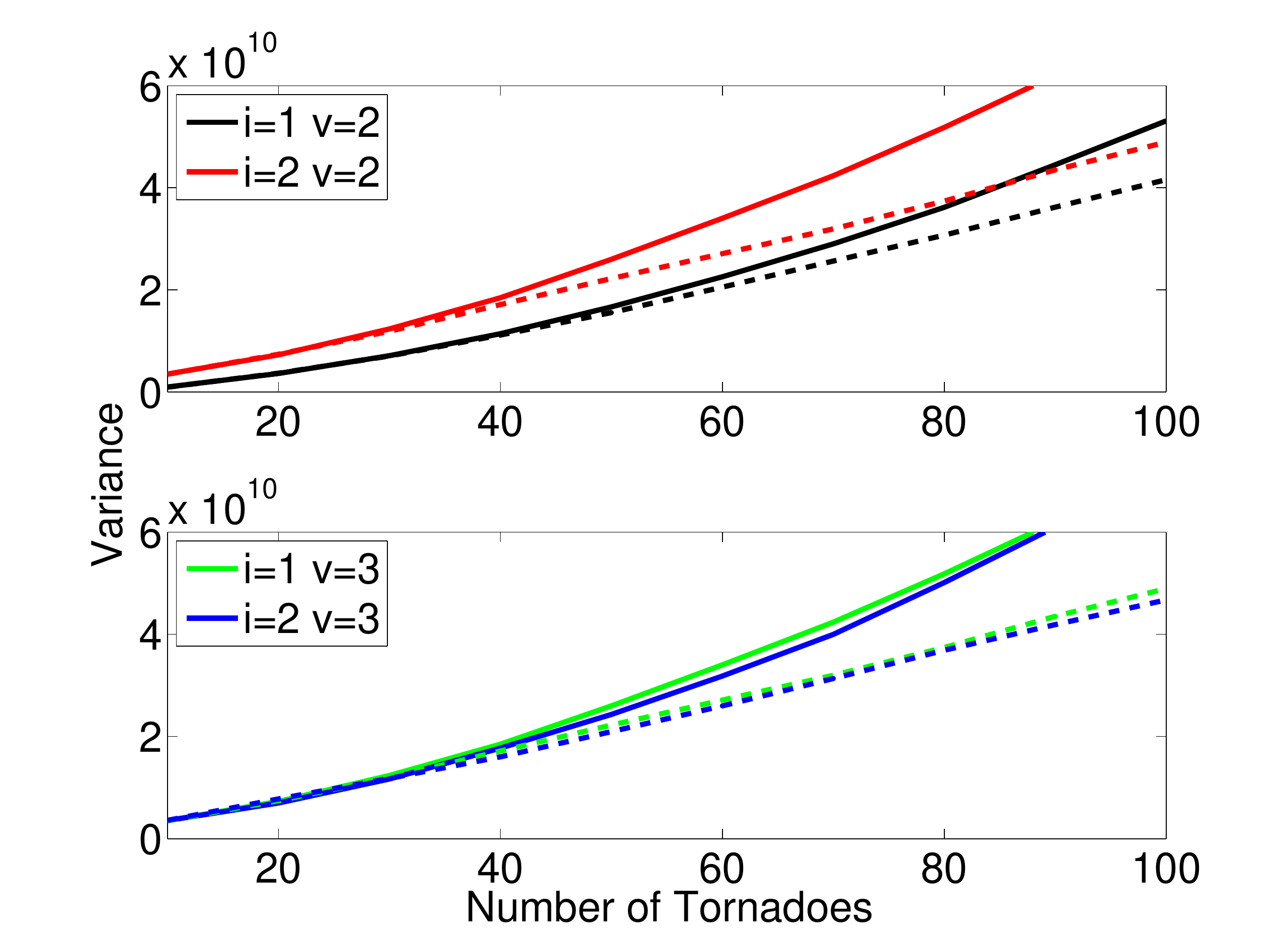}
	\caption{Variance of the accumulated discounted reward. Comparison between real (continuous line) and synthetic data (dashed line).}
	\label{var}
\end{figure}

As a further application of the proposed model we apply the reward model to the tornado time series. Particularly we transform the original process into costs that a state has to pay for the tornado damages. To do this we apply the results of \cite{simmons2013normalized}. The Fujita scale is then transformed into costs, so 8689$ \$ $, 62440$ \$ $, 121141$ \$ $, 146564$ \$ $, 177824$ \$ $ and 89192$ \$ $ that are respectively the mean costs of tornado degrees 0, 1, 2, 3, 4, 5. As previously said, we compute the expected value and the variance of the accumulated discounted reward, see Figure \ref{mean} and Figure \ref{var} respectively.  In both Figures the continuous lines are referred to real data while the dashed lines to the synthetic one. In these Figures we show the quantities as a function of the number of tornadoes and we highlight the dependences with the actual state $i$ and the backward process $v $ by varying them. It is possible to affirm that the semi-Markov model well caught the behaviors of the real data especially for the first number of tornadoes.

\section{Conclusion}

In this paper we model the statistical behaviors of tornadoes in a vast region of the USA. To do this we make use of a first order semi-Markov model that is more general of the Markov chain model. We show, through a statistical test that the latter one is not able to capture the duration effect of the tornadoes. The more general semi-Markov model in fact, by considering the time of permanence in a given state as generated by non memoryless distribution, is able to reproduce the duration effect.
Moreover, since we believe that the costs of the tornado damages are a serious problem related to this natural phenomenon, as an economic application we compute the expected value and the variance of the accumulated discounted cost and we show its dependency by the intensity and the duration of the initial tornado. 

\bibliographystyle{elsarticle-harv} 
\bibliography{b}

\end{document}